\documentclass[twocolumn,prb]{revtex4}
\usepackage{graphicx}
\usepackage{dcolumn}
\usepackage{bm}
\usepackage{amsmath}

\begin{document}

\preprint{}
\title{Controllable spin transport in ferromagnetic graphene junctions}
\author{Takehito Yokoyama}
\affiliation{Department of Applied Physics, Nagoya University, Nagoya, 464-8603, Japan}
\date{\today}

\begin{abstract}
We study spin transport in normal/ferromagnetic/normal graphene junctions where a gate electrode is attached to the ferromagnetic graphene. 
We find that due to the exchange field of the ferromagnetic graphene, spin current through the junctions has an oscillatory behavior with respect to the chemical potential in the ferromagnetic graphene, which can be tuned by the gate voltage. Especially, we obtain a controllable spin current reversal by the gate voltage. Our prediction of high controllability of spin transport in ferromagnetic graphene junction may contribute to the development of the spintronics. 
\end{abstract}

\maketitle



%

%



There is a rapidly growing attention to graphene because it has a rich potential from fundamental and applied physics point of view. \cite{Ando,Katsnelson,Castro}
Graphene has a two-dimensional honeycomb network of carbon atoms and, as a result, electrons in graphene are governed by Dirac equation, which provides a
 bridge between condensed matter physics and quantum electrodynamics, thus fascinating theoriticians toward the ultimate goal of the `unified' theory. \cite{Volovik,Tsvelik}
The progress of practical
fabrication techniques for single graphene sheets has
allowed experimental study of this system, which has attracted a
tremendous interest from the scientific community.\cite{novoselov,zhang,novoselov_nature}
There have been intensive studies on graphene to this date, for instance, half integer and unconventional quantum 
Hall effect\cite{zhang,Novoselov2,Yang}, observation of minimum conductivity\cite{novoselov_nature} and bipolar supercurrent\cite{heersche}.

Graphene has many important natures for applications: it exhibits gate-voltage-controlled carrier conduction, high field-effect mobilities and a small spin-orbit interaction.\cite{Kane,Hernando} Thus, it is extremely promising for future technologies, especially spintronics. 
Motivated by this, some studies on graphene have been implemented. 
 Spin injection into a graphene thin film has been successfully demonstrated by using non-local magnetoresistance measurements.\cite{Tombros,Ohishi,Cho} 
 The possibility of inducing ferromagnetic correlations in graphene due to the proximity effect by magnetic gates in close proximity to graphene is also discussed.\cite{Haugen} In these works, induced ferromagnetism in graphene is `extrinsic'.

On the other hand, there is another attempt to induce ferromagnetism `intrinsically' in graphene.  It has recently been predicted that zigzag edge graphene nanoribbon become half-metallic by an external transverse electric field due to the different chemical potential shift at the edges,\cite{Son,Kan,Dutta} which indicates the high controllability of ferromagnetism in graphene and hence opens the possibility of the spintronics application for graphene. In view of this, the study of spin transport in ferromagnetic graphene is very timely and desirable for the development of the spintronics. 

Stimulated by this, in this paper, we study spin transport in normal/ferromagnetic/normal graphene junctions where a gate electrode is attached to the ferromagnetic graphene. 
We find that due to the exchange field in the ferromagnetic graphene,  spin current through these systems has an oscillatory behavior with respect to the chemical potential in the ferromagnetic graphene, which can be tuned by the gate electrode.\cite{Novoselov2,zhang,novoselov_nature} Especially, we find a controllable spin current reversal by the gate voltage. Since the exchange field of graphene is also tunable by the inplane external electric field and even the half-metallicity can be induced\cite{Son,Kan,Dutta}, our prediction of high controllability of spin transport in ferromagnetic graphene junction may facilitate the development of the spintronics. Note that the model itself in this paper is the same as that in Ref.\cite{Haugen} and here we mainly focus on the case with the exchange field comparable to the Fermi energy in the model of Ref.\cite{Haugen}. 

Now, let us explain the formulation. 
The fermions around Fermi level in graphene obey a massless relativistic Dirac equation. 
The Hamiltonian is given by 
\begin{equation}
  H_\pm = v_F (\sigma_x k_x \pm \sigma_y k_y)
\end{equation}
with Pauli matrices $\sigma_x$ and $\sigma_y$, and the velocity $v_F\approx 10^{6}$ m/s in graphene. This is roughly 100 times larger than in a normal metal, and thus it is safe to neglect the Coulomb interaction compared to the kinetic energy in graphene. \cite{gonzales1999,kane2004}
The Pauli matrices operate on the two triangular sublattice space of the
honeycomb structure.  
The $\pm$ sign refers to the two so-called valleys of $K$ and
$K'$ points in the Brillouin zone. Also, there is a valley degeneracy, which
 allows one to consider one of the $H_\pm$ set. \cite{morpurgo2}
The linear dispersion relation is valid for Fermi levels 
as high as 1 eV,\cite{wallace} such that the fermions in graphene behave
like massless Dirac fermions in the low-energy regime.

\begin{figure}[htb]
\begin{center}
\scalebox{0.4}{
\includegraphics[width=19.0cm,clip]{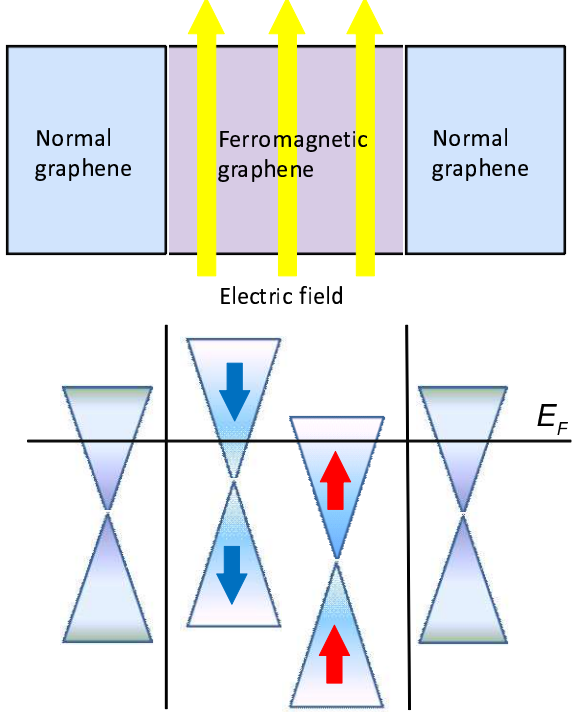}}
\end{center}
\caption{(color online) Schematic of the model of normal/ferromagnetic/normal graphene junction with corresponding Dirac cones. In the ferromagnetic graphene, different Dirac cones correspond to different chemical potentials of majority and minority spins.} \label{f1}
\end{figure}

We consider a two dimensional normal/ferromagnetic/normal graphene junction where a gate electrode is attached to the ferromagnetic graphene. This junction may be realized by applying an external transverse electric field to a part of graphene nanoribbon to make it ferromagnetic partially. 
  See Figure \ref{f1} for the schematic of the model with corresponding dispersion relations.  The interfaces are parallel to the $y$-axis and located at $x=0$ and $x=L$. Since there is a valley degeneracy, we focus on the Hamiltionian $H_+$ with $H_+ = v_F (\sigma_x k_x + \sigma_y k_y) - V(x)$,  $V(x)=E_F$ in the normal graphenes and $V(x)= E_F  + U \pm  H$ in the ferromagnetic graphene. Here, $E_F=v_F k_F$ is the Fermi energy, $U$ is the chemical potential shift tunable by the gate voltage, and $H$ is the exchange field. $\pm$ signs correspond to majority and minority spins. 
The wavefunctions are given by 
\begin{eqnarray}
 \psi_1 = \left( {\begin{array}{*{20}c}
   1  \\
   {e^{i\theta } }  \\
\end{array}} \right)e^{ip\cos \theta x +ip_y y} \nonumber \\
 + a_\pm \left( {\begin{array}{*{20}c}
   1  \\
   { - e^{ - i\theta } }  \\
\end{array}} \right)e^{ - ip\cos \theta x +ip_y y}  ,
\end{eqnarray}
\begin{eqnarray}
 \psi_2 = b_\pm \left( {\begin{array}{*{20}c}
   1  \\
   {e^{i\theta '} }  \\
\end{array}} \right)e^{ip'_\pm \cos \theta 'x +ip_y y} \nonumber \\
+ c_\pm \left( {\begin{array}{*{20}c}
   1  \\
   { - e^{ - i\theta '} }  \\
\end{array}} \right)e^{ - ip'_\pm \cos \theta 'x +ip_y y}  ,
\end{eqnarray}
\begin{eqnarray}
 \psi_3 = d_\pm \left( {\begin{array}{*{20}c}
   1  \\
   {e^{i\theta } }  \\
\end{array}} \right)e^{ip\cos \theta x +ip_y y} 
\end{eqnarray}
with angles of incidence $\theta$ and $\theta'$, $p = (E + E_F )/v_F$ and $ p'_\pm  = (E + E_F  + U \pm  H)/v_F$. 
Here, $\psi_1$ and $\psi_3$ are wavefunctions in the left and right normal graphenes, respectively, while $\psi_2$ is a wavefunction in the ferromagnetic graphene. 
Because of the translational symmetry in the $y$-direction, the momentum parallel to the $y$-axis is conserved: 
$p_y= p\sin \theta  = p'\sin \theta '$.

By matching the wave functions at the interfaces ($\psi_1=\psi_2$ at $x = 0$
and $\psi_2=\psi_3$ at $x = L$), we obtain the coefficients in the wavefunctions. 
Note that these conditions are equivalent to $\hat v_x \psi_1=\hat v_x \psi_2$ at $x = 0$
and $\hat v_x \psi_2=\hat v_x \psi_3$ at $x = L$ with velocity operator $\hat v_x  = \partial H_ +  /\partial k_x  = v_F\sigma _x$ and hence the current is conserved at the interfaces. 
The transmittion coefficient has the form 
\begin{widetext}
\begin{eqnarray}
d_\pm = \frac{{\cos \theta \cos \theta '{\mathop{\rm e}\nolimits} ^{ - ipL\cos \theta } }}{{\cos(p'_\pm L\cos \theta ')\cos \theta \cos \theta ' - i\sin (p'_\pm L\cos \theta ')(1 - \sin \theta \sin \theta ')}}.
\end{eqnarray}
\end{widetext}

Then, the dimensionless spin-resolved conductances $G_{\uparrow,\downarrow}$ are given by 
\begin{eqnarray}
G_{\uparrow,\downarrow}   = \frac{1}{2}\int_{ - \pi /2}^{\pi /2} {d\theta \cos \theta T_{\uparrow,\downarrow}  (\theta )} 
\end{eqnarray}
with $T_{\uparrow,\downarrow} (\theta ) = \left| {d_\pm(\theta )} \right|^2$. 
Finally, the spin conductance $G_s$ is defined as 
$G_s  = G_ \uparrow   - G_ \downarrow$. Below, we focus on the zero voltage conductances, namely we set $E=0$.

Now, let us explain the underlying mechanism of spin manipulation by the gate voltage. 
In the limit of $\left| U \pm H \right| \gg E_F$, we have $\theta' \to 0$ and hence the transmittion coefficient of the form 
\begin{eqnarray}
d_\pm \to \frac{{\cos \theta {\mathop{\rm e}\nolimits} ^{ - ipL\cos \theta } }}{{\cos\chi_\pm  \cos \theta  - i\sin \chi_\pm  }}
\end{eqnarray}
with $\chi _\pm   = \chi  \pm \chi _H, \chi  = UL/v_F,$ and $\chi _H  =HL/v_F$. 
The resulting transmittion probability is represented as \cite{katsnelson}
\begin{eqnarray}
T_{ \uparrow , \downarrow } (\theta ) \to \frac{{\cos ^2 \theta }}{{1 - \sin ^2 \theta \cos^2 \chi _\pm }}.
\end{eqnarray}
From this expression, we see the $\pi$-periodicity with respect to $\chi_\pm$ or $\chi$. \cite{Haugen,katsnelson,linder,sengupta}
 We also find that $G_{\uparrow,\downarrow} $ has a maximum (minimum) value of 1 (2/3) at $\chi_\pm=0$ $(\pi/2)$. 
The phase difference between $G_{\uparrow}$ and $G_{\downarrow} $ is given by $\chi _+ - \chi_- = 2\chi _H  = 2HL/v_F$. If this is equal to the half period $\pi/2$ (namely, $H/E_F=\pi k_F L/4$), we expect a large spin current which oscillates with $\chi$, namely the gate voltage, because when one of $G_{\uparrow}$ and $G_{\downarrow} $ has a maximum at a certain $\chi$, the other has a minimum at the same $\chi$. In this case, the value of $G_s$ oscillates between $-1/3$ and $1/3$. Note that the electrical conductance $G_{\uparrow}+G_{\downarrow} $ in the junctions is always positive and therefore spin current reversal in our model is not accompanied with the current reversal.

\begin{figure}[htb]
\begin{center}
\scalebox{0.4}{
\includegraphics[width=19.0cm,clip]{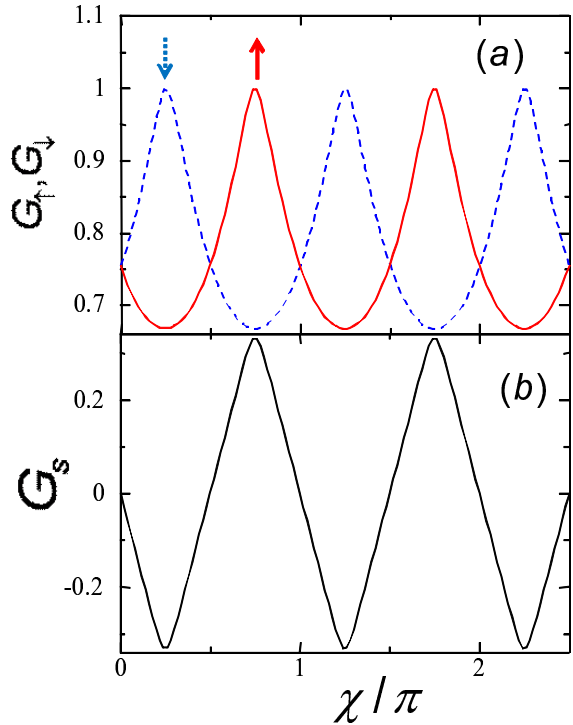}}
\end{center}
\caption{(color online) Plots as a function of $\chi$ tunable by  the gate voltage, in the limit of $U \to \infty$. (a) spin resolved conductances where the phases of 
$G_ \uparrow$ (solid line) and $G_ \downarrow$ (dotted line) are shifted by half period, $\chi _\uparrow - \chi _\downarrow= \pi/2$. (b) Spin conductance $G_s$ which oscillates with the period $\pi$ with respect to $\chi$ but is never damped. } \label{f2}
\end{figure}

\begin{figure}[htb]
\begin{center}
\scalebox{0.4}{
\includegraphics[width=20.0cm,clip]{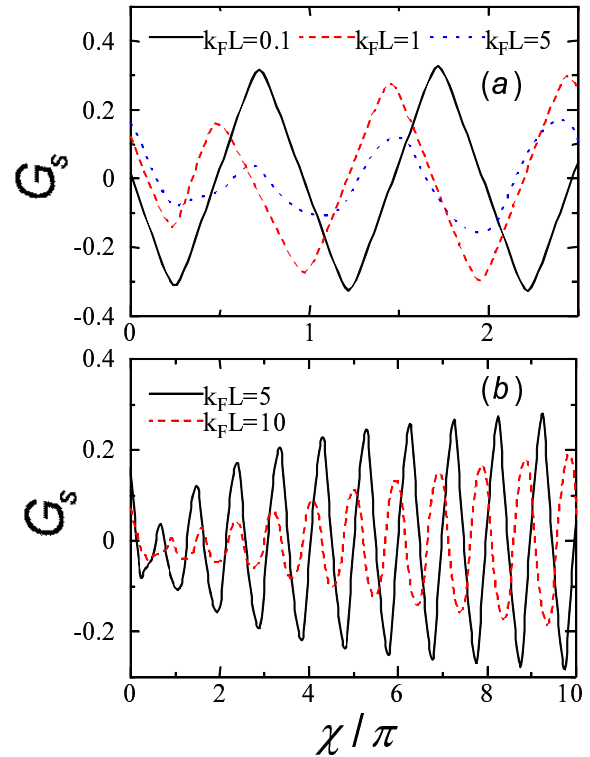}}
\end{center}
\caption{(color online). Spin conductance $G_s$ as a function of $\chi$ for several values of $k_F L$ with $\chi _+ - \chi _-= \pi/2$.} \label{f3}
\end{figure}

\begin{figure}[htb]
\begin{center}
\scalebox{0.4}{
\includegraphics[width=20.0cm,clip]{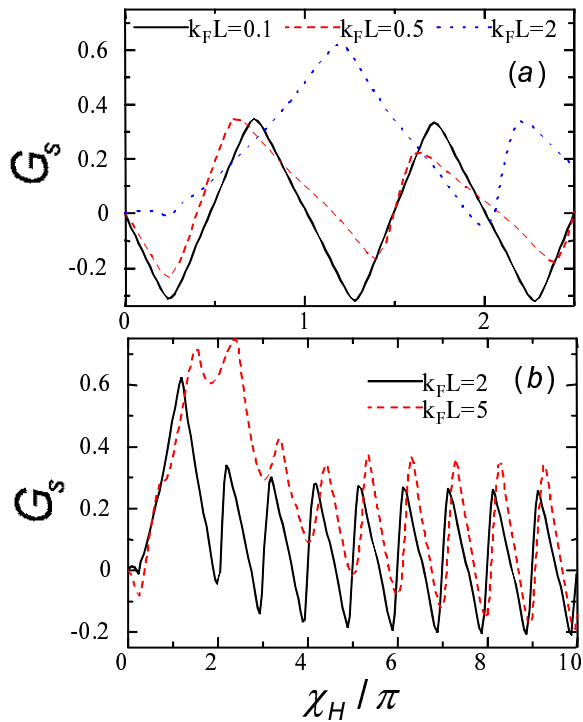}}
\end{center}
\caption{(color online) $G_s$ as a function of $\chi_H$, which is proportional to  the exchange field, for several values of $k_F L$. Here, we choose $\chi=\pi/4$. } \label{f4}
\end{figure}

The results in this limiting case are shown in Fig. \ref{f2}.
 Figure \ref{f2} (a) shows spin resolved conductances as a function of $\chi$ which is tunable by the gate voltage. Here, the phases of $G_ \uparrow$ and $G_ \downarrow$ are shifted by half period, $\chi _+ - \chi _-= \pi/2$. 
Then, we have a finite spin current as shown in  Fig. \ref{f2} (b). Remarkably, it oscillates with the period $\pi$ with respect to $\chi$ but is never damped. As is seen, we can reverse the spin current by changing the gate voltage. Here, we focus on the limit of $\left| U \pm H \right| \gg E_F$. Next, we consider more general cases without taking this limit.

 Figure \ref{f3} (a) exhibits the spin conductance $G_s$ as a function of $\chi$ for several values of $k_F L$. For a ferromagnetic graphene with $k_F L=0.1$, the $\pi$-periodicity is seen. With increasing $k_F L$, however, the $\pi$-periodicity gets broken as seen in Fig. \ref{f3} (a). This is because $U \gg E_F$ is no more satisfied for small $k_F L$ since $\chi  = k_F L U/E_F$.  
Thus, by choosing large $\chi$ for large $k_F L$, we again get $\pi$-periodicity of  $G_s$ with respect to $\chi$ as shown in Fig. \ref{f3} (b). From this figure, we find that $U/E_F > 1$ is required to obtain controllable spin current reversal. 

Figure \ref{f4} displays $G_s$ as a function of $\chi_H$, which is proportional to  the exchange field, for several values of $k_F L$. Here, we fix $\chi=\pi/4$ to obtain finite spin current.  For a short ferromagnetic graphene with $k_F L=0.1$, the $\pi$-periodicity is satisfied.  With increasing $k_F L$, the $\pi$-periodicity requires larger magnitude of $\chi_H$ as shown in this figure, similar to Fig. \ref{f3}. 

Our prediction is applicable as long as the continuum Dirac equation is valid, which requires wide graphene nanoribbon. 
For the realization of our prediction of controllable spin current by the gate voltage, $U/E_F > 1$ is required. If we choose ferromagnetic graphene with  $ k_F L =1 $ and $E_F \simeq$ 1 meV as an example, we need the length around 1 $\mu$m. Also, $\chi \sim U/E_F, \chi_H \sim H/E_F$ and hence $U, H \sim$ 1-10 meV is required. These values can be achieved by the present experimental technique. \cite{Novoselov2,zhang,novoselov_nature,Son,Kan,Dutta}

In summary, we studied spin transport in normal/ferromagnetic/normal graphene junctions where a gate electrode is attached to the ferromagnetic graphene. 
We found that because of the exchange field in the ferromagnetic graphene,  spin current through the junctions has an oscillatory behavior with respect to the chemical potential in the ferromagnetic graphene, which can be tuned by the gate electrode.\cite{Novoselov2,zhang,novoselov_nature} Especially, spin current reversal by the gate voltage, which is not accompanied with the current reversal, is obtained. The exchange field of graphene is also known to be tunable by the inplane external electric field.\cite{Son,Kan,Dutta}  Therefore, our prediction of high controllability of spin transport in ferromagnetic graphene junctions will contribute to the development of the spintronics.


The author acknowledges support by the JSPS. 
%



\begin{thebibliography}{99}
\bibitem{Ando} T. Ando, J. Phys. Soc. Jpn. \textbf{74}, 777 (2005).

\bibitem{Katsnelson} M. I. Katsnelson and K. S. Novoselov, Solid State Commun. \textbf{143}, 3 (2007).

\bibitem{Castro} A. H. Castro Neto, F. Guinea, N. M. R. Peres, K. S. Novoselov  and A. K. Geim, arXiv:0709.1163v1.


\bibitem{Volovik} G. E. Volovik, \textit{ The Universe in a Helium Droplet} (Clarendon, Oxford, 2003).

\bibitem{Tsvelik} A. M. Tsvelik, \textit{Quantum Field Theory in Condensed Matter Physics} (Cambridge Univ Pr, 2007).

\bibitem{novoselov}  K. S. Novoselov, A. K. Geim, S. V. Morozov, D. Jiang, Y. Zhang, S. V. Dubonos, I. V. Grigorieva and A. A. Firsov, Science \textbf{306}, 666 (2004).

\bibitem{zhang} Y. Zhang, Y.-W. Tan, H. L. Stormer, and P. Kim, Nature \textbf{438}, 201 (2005).

\bibitem{novoselov_nature} K. S. Novoselov, A. K. Geim, S. V. Morozov, D. Jiang, M. I. Katsnelson, I. V. Grigorieva, S. V. Dubonos, and A. A. Firsov, Nature \textbf{438}, 197 (2005).

\bibitem{Novoselov2} K. S. Novoselov, E. McCann, S. V. Morozov, V. I. Fal'ko, M. I. Kastenelson, U. Zeitler, D. Jiang, F. Schedin, and A. K. Geim, Nat. Phys. \textbf{2}, 177 (2006).

\bibitem{Yang} K. Yang, Solid State Commun. \textbf{143}, 27 (2007).

\bibitem{heersche}  H. B. Heersche, P. Jarillo-Herrero, J. B. Oostinga, L. M. K. Vandersypen, and A. F. Morpurgo, Nature \textbf{446}, 56 (2007).




\bibitem{Kane} C. L. Kane and E. J. Mele, Phys. Rev. Lett. \textbf{95}, 226801 (2005). 

\bibitem{Hernando} D. Huertas-Hernando, F. Guinea, and A. Brataas, Phys. Rev. B \textbf{74}, 155426 (2006).

\bibitem{Tombros} N. Tombros, C. Jozsa, M. Popinciuc, H. T. Jonkman, and B. J. van Wees, Nature (London) \textbf{448}, 571 (2007). 




\bibitem{Ohishi} M. Ohishi, M. Shiraishi, R. Nouchi, T. Nozaki, T. Shinjo and Y. Suzuki, Jap. J. Appl. Phys. \textbf{46}, L605 (2007).

\bibitem{Cho} S. Cho, Yung-Fu Chen, and M. S. Fuhrer, Appl. Phys. Lett. \textbf{91}, 123105 (2007). 

\bibitem{Haugen} H. Haugen, Daniel Huertas-Hernando, and Arne Brataas, arXiv:0707.3976v1.

\bibitem{Son} Y.-W. Son, M. Cohen, and S. G. Louie, Nature (London) \textbf{444}, 347 (2006).

\bibitem{Kan} Er-Jun Kan, Zhenyu Li, Jinlong Yang, and J. G. Hou, arXiv:0708.1213v1.

\bibitem{Dutta} S. Dutta and S. K. Pati, arXiv:0712.4130v1. 

\bibitem{gonzales1999} J. Gonz\'alez, F. Guinea, M. A. H. Vozmediano, Phys. Rev. B \textbf{59}, R2474 (1999).

\bibitem{kane2004} C. L. Kane and E. J. Mele, Phys. Rev. Lett. \textbf{93}, 197402 (2004).

\bibitem{morpurgo2} A. F. Morpurgo and F. Guinea, Phys. Rev. Lett. \textbf{97}, 196804 (2006).

\bibitem{wallace} P. R. Wallace, Phys. Rev. \textbf{71}, 622 (1947).

\bibitem{katsnelson} M. I. Katsnelson, K. S. Novoselov, and A. K. Geim, Nature Phys. \textbf{2}, 620 (2006).

\bibitem{sengupta} S. Bhattacharjee and K. Sengupta, Phys. Rev. Lett. \textbf{97}, 217001 (2006).

\bibitem{linder} J. Linder and A. Sudb{\o}, Phys. Rev. Lett. \textbf{99}, 147001 (2007).





\end{thebibliography}
\end{document}